\documentclass[pdflatex,sn-mathphys-num]{sn-jnl}

\usepackage{graphicx}
\usepackage{multirow}
\usepackage{amsmath,amssymb,amsfonts}
\usepackage{amsthm}
\usepackage{mathrsfs}
\usepackage[title]{appendix}
\usepackage{xcolor}
\usepackage{textcomp}
\usepackage{manyfoot}
\usepackage{booktabs}
\usepackage{algorithm}
\usepackage{algorithmicx}
\usepackage{algpseudocode}
\usepackage{listings}
\usepackage{tikz}
\usetikzlibrary{patterns}
\usepackage{pgfplots}
\pgfplotsset{compat=1.18}
\usepackage{bookmark}
\hypersetup{hypertexnames=false}

\theoremstyle{thmstyleone}

\theoremstyle{thmstyletwo}

\theoremstyle{thmstylethree}

\begin{document}

\title[Cross-Language AI Reputation Divergence]{The Language Blind Spot: How Query Language and Brand Recognition Tier Shape AI-Constructed Brand Reputation Across Twelve European Languages}

\author*[1,2]{\fnm{Dmitrij} \sur{\.Zatuchin}}\email{dmitrij.zatuchin@eek.ee}

\affil*[1]{\orgdiv{Department of Information Technologies},
  \orgname{Estonian Entrepreneurship University of Applied Sciences (EUAS)},
  \orgaddress{\city{Tallinn}, \country{Estonia}}}
\affil[2]{\orgname{Rankfor.AI},
  \orgaddress{\city{Tallinn}, \country{Estonia}}}

\abstract{Large language models (LLMs) increasingly mediate how consumers and professionals form impressions of organisations. Most monitoring of this mediation is conducted in English, on the assumption that an English-language query returns a representative picture. This study measures how far that assumption holds. We queried three grounded LLMs (GPT-5.4, Google Gemini 3.1 Pro, and Perplexity Sonar Pro) about 66 brands drawn from eleven Northern, Baltic, and Central European home markets, in twelve languages spanning four language families (Germanic, Uralic, Baltic, Slavic), generating 35{,}640 grounded responses. Multilingual embeddings (BGE-M3) allow direct cross-language comparison without translation. Three results emerge within this sample. First, AI-constructed reputation is language-bound: mean cross-language cosine similarity is 0.825, responses in the same language family are more similar than across families (0.844 vs.\ 0.820; $t = 57.98$, $p < 0.001$, Cohen's $d = 0.31$), and sentiment varies systematically by language ($F = 268.5$, $p < 0.001$, $\eta^2 = 0.077$), with Uralic and Baltic languages the most positive and Germanic languages, including English, the most critical. Hierarchical clustering of the twelve languages recovers the Slavic and Baltic families as clean groups (cophenetic correlation 0.915). Second, query language shifts \emph{which} brands are recommended far more than \emph{how} they are described. Moving from an English query to a brand's home language raises recommendation share by 0.80 (on a 0--1 scale) for local champions but only 0.15 for global multinationals ($t = -8.84$, $p < 0.001$); the same shift produces no comparable reversal in sentiment. An English-only audit therefore understates a local champion's AI visibility while representing a multinational's fairly. Third, response \emph{stability} varies more with model choice than with language: across a five-iteration replication on a 20-brand subset, model differences dominate language differences ($\eta^2_{\text{model}} = 0.32$ vs.\ $\eta^2_{\text{language}} = 0.01$). A re-derivation of an earlier exploratory claim that AI brand visibility is two-dimensional finds two interpretable axes that account for 60.8\% of variance but a third weaker dimension at this larger scale, so we treat dimensionality as approximate. These results indicate that English-only AI reputation monitoring leaves a measurable language blind spot, concentrated in the visibility of locally headquartered brands.}

\keywords{Large language models, Cross-language AI, Brand reputation, Multilingual NLP, Generative engine optimisation, Recommendation share, AI bias, Sentiment analysis}

\maketitle

\section{Introduction}\label{sec:intro}

Large language models (LLMs) now sit between organisations and their stakeholders as a layer of reputation mediation \cite{fombrun1990whatsnamereputation}. When a buyer asks an AI assistant which logistics provider to trust or which fintech to use, the model's answer draws on an opaque mixture of parametric knowledge and retrieved sources, and the resulting narrative shapes the buyer's shortlist. Prior work has shown that this mediation is neither neutral nor stable: recommendations carry systematic gender disparities \cite{zatuchin2026a}, and the brand that a model names as the category leader changes from one model to the next \cite{zatuchin2026c}.

One dimension of this mediation has received little measurement: the language of the query. Most brands and most monitoring tools audit AI visibility in English. If the model returns a representative picture regardless of query language, English-only monitoring is sufficient. If it does not, then a marketing team in Vilnius or Helsinki reading English-language AI outputs may be looking at a different brand than the one their customers encounter in Lithuanian or Finnish.

There is reason to expect divergence. Pre-training corpora are dominated by English; the smaller Northern, Baltic, and Central European languages each occupy a thin slice of the data \cite{dodge2021documenting, conneau2020unsupervised}, and NLP quality tracks the volume of language-specific data \cite{blasi2022systematic}. Parametric knowledge about specific entities is distributed unevenly across languages and entity popularity \cite{kandpal2022large, mallen2023trust}. Whether these asymmetries change the \emph{content and valence} of what a model says about a fixed real-world entity, holding the entity constant and varying only the query language, is an empirical question this study addresses at scale.

We study 66 brands across eleven home markets (Lithuania, Latvia, Estonia, Finland, Sweden, Norway, Denmark, Germany, Poland, Czech Republic, Slovakia), queried in twelve languages drawn from four language families, on three grounded models, producing 35{,}640 responses. The design extends a line of empirical work on AI-mediated brand reputation \cite{zatuchin2026c} in which cross-linguistic variation remained the primary unaddressed dimension. A 20-brand companion cohort, collected under an identical protocol with a five-iteration stability extension, supplies the replication and stability measurements that the main corpus does not.

We organise the analysis around two questions, each with measurable sub-parts:

\begin{description}
    \item[Q1 (What is said).] Does the content and valence of AI-constructed reputation depend on query language, and is the dependence structured by language family?
    \begin{description}
        \item[Q1a] Mean cross-language semantic similarity, and same-family versus cross-family difference.
        \item[Q1b] Sentiment by language and by language family.
        \item[Q1c] Whether the home-language effect differs between locally headquartered champions and global multinationals, and whether it acts on \emph{how} brands are described or on \emph{which} brands are recommended.
        \item[Q1d] Whether clustering languages by AI-narrative similarity recovers language-family structure.
    \end{description}
    \item[Q2 (How stable it is).] Is response stability under repeated querying moderated more by query language or by model choice?
\end{description}

We frame the study as measurement. We map how AI-constructed reputation varies across languages within a defined sample of Northern and Central European brands and models, report effect sizes with their uncertainty, and are explicit about what the measurements do and do not license. We do not claim that AI-constructed reputation tracks human reputation; we measure a property of the models' outputs, which is itself consequential because those outputs are what users read.

\section{Related Work}\label{sec:related}

\subsection{AI as Information Intermediary}

The role of media intermediaries in shaping which entities receive public attention is long established in agenda-setting theory \cite{mccombs1972agenda}. Applied to LLMs, the framework casts models as algorithmic gatekeepers whose linguistic capabilities determine which information reaches users in each language market. The corporate-reputation literature treats reputations as socially constructed from observable signals \cite{fombrun1990whatsnamereputation, vanriel1997managing}, operationalised through multi-stakeholder, multi-market survey systems such as RepTrak \cite{fombrun2015reptrak}. The language through which AI mediates those signals is a previously under-measured part of that construction.

Recent empirical work characterises LLM gatekeeping behaviour directly. \.Zatuchin \cite{zatuchin2026a} documented gender disparities in brand recommendations. \.Zatuchin \cite{zatuchin2026c} mapped competitive category ownership across 50 brands and reported that the three studied models agreed on the top-recommended brand in only 41.6\% of categories. These studies show that AI-mediated reputation is variable, platform-dependent, and subject to systematic disparities. None measured query language as a source of variation across a large multilingual sample.

\subsection{Multilingual AI and Language Disparity}

The uneven distribution of languages in LLM training data is well documented. Dodge et al.\ \cite{dodge2021documenting} found English dominates the C4 corpus, with most other languages below one percent; curated multilingual corpora retain a heavy English skew \cite{conneau2020unsupervised}. The imbalance has measurable consequences for quality: identical-question benchmarks across dozens of languages show large accuracy drops in the lowest-resource ones \cite{blasi2022systematic, xuan2025mmluprox, sayeedi2025translate}, and parametric factual knowledge is distributed unevenly across languages and entity popularity \cite{kandpal2022large, mallen2023trust, petroni2019language}. For the languages in this study the resource landscape is mixed: German is among the best-resourced web languages; Swedish, Danish, Norwegian, Polish, and Czech are mid-resource; Finnish, Estonian, Latvian, Lithuanian, and Slovak are comparatively low-resource \cite{kaalep2001estonian}. This spread, crossing four language families, lets us separate two candidate mechanisms: divergence driven by structural language distance, which would track family boundaries, and divergence driven by training-data volume, which would track resource level regardless of family. Recent work shows the dependence runs deeper than benchmark accuracy: multilingual models store the same fact inconsistently across languages \cite{wang2025lost}, low-frequency non-English facts often surface only through their English equivalents \cite{liu2025tracing}, and production assistants frequently route non-English queries through English retrieval \cite{peec2026english}. An English baseline is therefore a noisy reference, not a neutral one, a point we treat as a measurement caveat.

Prior multilingual-LLM research has concentrated on task performance, such as translation quality \cite{costa2022no}, cross-lingual benchmark accuracy \cite{ahuja2023mega}, and transfer learning \cite{pires2019multilingual}, while the fairness literature has concentrated on social biases encoded in training data \cite{bolukbasi2016man, bender2021stochastic}. Measuring whether the evaluative content of LLM outputs about fixed real-world entities varies across languages is a distinct and largely open question.

\subsection{Corporate Reputation Across Cultures}

Corporate reputation is culturally situated. Gardberg and Fombrun \cite{gardberg2006corporate} showed that reputation dimensions vary in salience across national contexts, and Deephouse et al.\ \cite{deephouse2016corporate} demonstrated that media environments and stakeholder expectations shape reputation differently across markets. If LLMs encode these cultural differences, cross-language variation in AI outputs may partly reflect genuine cross-cultural patterns rather than only training-data artefacts. The present study measures the variation; it does not adjudicate the mechanism, which would require human-perception baselines per market that we discuss as a limitation.

\section{Methodology}\label{sec:methods}

\subsection{Design}

The main corpus uses a fully crossed design over four factors: language (12 levels), model (3 levels), brand (66 levels, nested within eleven home markets and thirteen industries), and prompt (15 templates across five categories). A 20-brand companion cohort (the Central and Eastern European, or CEE, cohort) was collected under the same prompt protocol and additionally repeated each stability probe across five iterations, supplying the replication and stability measurements reported in Section~\ref{sec:stability}.

\subsection{Brand and Market Selection}

Sixty-six brands were selected, six per market across eleven home markets, balancing globally recognised multinationals (for example Volkswagen, SAP, IKEA, Spotify, Maersk, Novo Nordisk) with nationally or regionally significant champions (for example Lietuvos pa\v{s}tas, airBaltic, Selver, Vipps, Alza.cz, Tatra Banka). Each brand was hand-classified by recognition tier (global, pan-European, Nordic-Baltic, national); the classification is used in Section~\ref{sec:tierflip} and its subjectivity is noted as a limitation. The full brand roster, with market, industry, and tier, is provided in the supplementary materials.

\subsection{Languages}

Twelve languages were queried, spanning four families and a wide resource range (Table~\ref{tab:languages}). English serves as the cross-language baseline against which home-language outputs are compared.

\begin{table}[t]
\caption{Language sample: 12 languages across 4 families. Resource level is an approximate ranking of language-specific web and training-corpus volume.}\label{tab:languages}
\centering
\small
\begin{tabular}{lllc}
\toprule
\textbf{Family} & \textbf{Languages (code)} & \textbf{Resource level} & \textbf{Home markets} \\
\midrule
Germanic & English (en), German (de), Swedish (sv), & en/de excellent; & DE, SE, NO, DK \\
         & Norwegian (no), Danish (da) & sv/no/da good & (en baseline) \\
Uralic   & Finnish (fi), Estonian (et) & moderate / limited & FI, EE \\
Baltic   & Lithuanian (lt), Latvian (lv) & limited & LT, LV \\
Slavic   & Polish (pl), Czech (cs), Slovak (sk) & moderate / moderate / limited & PL, CZ, SK \\
\bottomrule
\end{tabular}
\end{table}

\subsection{Prompt Protocol}

Fifteen prompt templates were used across five categories: open reputation (A), source elicitation (B), competitive positioning (C), regional context (E), and visibility/media (F), three templates each. A sixth category of stability probes (D) was collected on the companion cohort with five iterations each. All templates were translated into each target language and verified for semantic equivalence by back-translation. Two of the categories (E regional context, F visibility) contain buyer-intent prompts of the form ``Who are the leading \{industry\} companies in \{market\}?''; these support the recommendation-share measure in Section~\ref{sec:tierflip}.

\subsection{Models}

Three grounded models were queried through their official APIs: GPT-5.4 (OpenAI), Gemini 3.1 Pro (Google), and Perplexity Sonar Pro (Perplexity). All three perform web grounding, so responses reflect retrieved sources as well as parametric knowledge. Configuration was held constant: temperature $0.1$, maximum output 4{,}096--8{,}192 tokens, with a language-aware system instruction directing the model to answer in the target language. Gemini's grounding returns vertex redirector URLs; these were resolved to publisher domains via the citation title before source classification, recovering roughly 23{,}000 citations that a naive parse would drop.

\subsection{Data Collection}

The main corpus comprises $15 \times 3 \times 12 \times 66 = 35{,}640$ responses, collected April--May 2026 with automatic checkpointing and resume. No responses were lost to API errors (0\% error rate after retries). The companion cohort adds the five-iteration stability probes used in Section~\ref{sec:stability}.

\subsection{Analytical Methods}

\paragraph{Multilingual embeddings.} All responses were embedded with BGE-M3 \cite{chen2024bge}, a multilingual model producing L2-normalised 1{,}024-dimensional vectors that place texts from different languages in a shared space, so cosine similarity compares meaning across languages without translation.

\paragraph{Sentiment.} Sentiment was scored with the multilingual XLM-RoBERTa model \texttt{cardiffnlp/twitter-xlm-roberta-base-sentiment} on a $[-1, +1]$ scale. The model is known to read long, structured, list-style responses as more negative than a human would; we therefore treat absolute polarity as noisy and rely on relative cross-language patterns and on the translation-free embedding similarity, which does not depend on the sentiment model. The calibration issue is revisited in the limitations.

\paragraph{Recommendation share.} For the buyer-intent prompts, recommendation share is the proportion of responses in a (brand $\times$ language) cell in which the brand is named, on a $[0, 1]$ scale. This is a transparent, model-free count and is the primary measure for the tier analysis.

\paragraph{Source attribution.} Citations were classified into eleven source types (tier-1 news, academic, government, industry report, review platform, financial directory, company-owned, Wikipedia, social media, other web, implicit knowledge) by URL-domain matching with a keyword fallback. The main corpus yields 150{,}093 classified citations across 21{,}077 unique domains.

\paragraph{Stability.} On the companion cohort, stability is the mean pairwise cosine similarity across the five iterations of each (brand $\times$ prompt $\times$ model $\times$ language) cell, in $[0, 1]$.

\paragraph{Statistical tests.} Cross-language similarity is summarised by same-family versus cross-family Welch $t$-tests with Cohen's $d$. Sentiment and stability are tested by one-way ANOVA with $\eta^2$ effect sizes computed for language and for model separately. The source-type by language association uses a chi-square test with Cram\'{e}r's $V$. Language clustering uses average-linkage hierarchical clustering on $1 - $ mean cosine, with cophenetic correlation as dendrogram fidelity. The recognition-tier contrast uses a Welch $t$-test on the home-minus-English delta. The dimensionality re-derivation (Section~\ref{sec:dimensionality}) uses standardised principal component analysis on brand-level measures.

\section{Results}\label{sec:results}

\subsection{Corpus Overview}

The analysed corpus is 35{,}640 responses with no errors, evenly distributed at 2{,}970 responses per language. Grounded responses are long: median length is 289 words in English, falling to 189 in Finnish, with model differences larger than language differences (Gemini median 411 words, OpenAI 190, Perplexity 178). As in prior work, longer English responses provide more surface for both negative phrasing and source extraction; we note response length as a confound where relevant.

\subsection{Q1a: Cross-Language Semantic Divergence}\label{sec:semantic}

Across 196{,}020 cross-language response pairs, mean cosine similarity is 0.825. The divergence is structured by language family: same-family pairs average 0.844 and cross-family pairs 0.820 (Welch $t = 57.98$, $p < 0.001$, Cohen's $d = 0.31$, small-to-medium). AI narratives about the same brand are measurably more alike when the two queries share a language family than when they cross one. The effect is modest in magnitude, which is itself informative: family membership shapes the narrative but does not partition it into disconnected stories.

\subsection{Q1b: Sentiment by Language and Family}\label{sec:sentiment}

\begin{figure}[t]
\centering
\includegraphics[width=\textwidth]{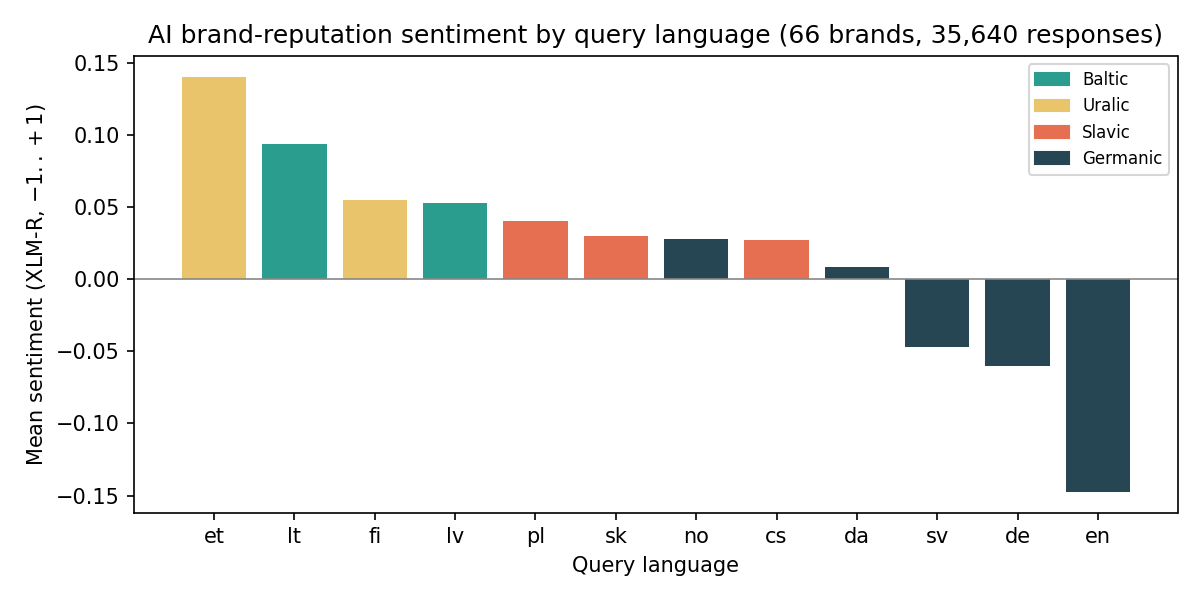}
\caption{Mean AI brand-reputation sentiment by query language, coloured by language family. Uralic and Baltic languages read most positive; Germanic languages, including English, read most critical.}\label{fig:sentiment}
\end{figure}

Sentiment varies systematically across languages (one-way ANOVA $F = 268.5$, $p < 0.001$, $\eta^2 = 0.077$). Table~\ref{tab:sentiment} and Figure~\ref{fig:sentiment} show the pattern. Estonian and Lithuanian produce the most positive readings; English and German the most critical. Aggregated by family, the ordering is Uralic ($+0.098$) $>$ Baltic ($+0.074$) $>$ Slavic ($+0.033$) $>$ Germanic ($-0.044$).

\begin{table}[t]
\caption{Mean sentiment by query language (XLM-R, $[-1,+1]$), 2{,}970 responses each. Family in parentheses. Absolute polarity is noisy (see Methods); the cross-language ordering is the interpretable signal.}\label{tab:sentiment}
\centering
\small
\begin{tabular}{lc lc}
\toprule
\textbf{Language} & \textbf{Mean} & \textbf{Language} & \textbf{Mean} \\
\midrule
Estonian (Ura) & $+0.140$ & Slovak (Sla) & $+0.030$ \\
Lithuanian (Bal) & $+0.094$ & Norwegian (Ger) & $+0.027$ \\
Finnish (Ura) & $+0.055$ & Czech (Sla) & $+0.027$ \\
Latvian (Bal) & $+0.053$ & Danish (Ger) & $+0.009$ \\
Polish (Sla) & $+0.040$ & Swedish (Ger) & $-0.047$ \\
 & & German (Ger) & $-0.060$ \\
 & & English (Ger) & $-0.147$ \\
\bottomrule
\end{tabular}
\end{table}

The pattern runs opposite to a naive ``low-resource means worse'' expectation: the lower-resourced Uralic and Baltic languages return the most positive sentiment, while the best-resourced Germanic languages return the most critical. A plausible reading, which we cannot confirm without a media-volume covariate, is that thicker native-language coverage in Germanic markets surfaces more critical journalism into the model's grounding, while thinner coverage in smaller languages leaves a more neutral or promotional baseline. For sentiment, language explains more variance than model ($\eta^2_{\text{language}} = 0.077$ vs.\ $\eta^2_{\text{model}} = 0.023$); the model means are close (OpenAI $+0.074$, Perplexity $-0.003$, Gemini $-0.016$).

\subsection{Q1c: The Tier-Flip Acts on Recommendation More Than Sentiment}\label{sec:tierflip}

\begin{figure}[t]
\centering
\includegraphics[width=0.92\textwidth]{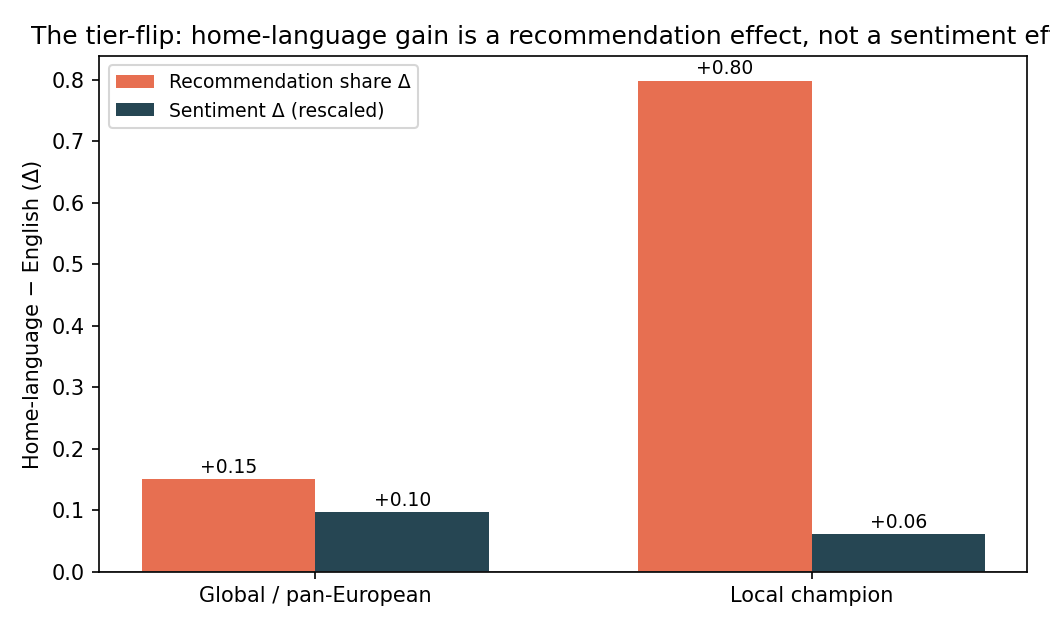}
\caption{Home-language minus English change for local champions versus global brands. The home-language advantage is large for recommendation share and concentrated in local champions; the sentiment change is small and does not separate the tiers.}\label{fig:tierflip}
\end{figure}

The most consequential pattern concerns visibility rather than tone. We compared each brand's home-language outputs to its English outputs and split the 65 non-English-home brands by recognition tier (24 local champions: Nordic-Baltic or national; 41 global or pan-European). The contrast is reported on two transparent measures and the composite for corroboration (Figure~\ref{fig:tierflip}).

On \textbf{recommendation share}, moving from English to the home language raises a local champion's share by $+0.80$ on the $[0,1]$ scale, against $+0.15$ for a global brand (Welch $t = -8.84$, $p < 0.001$). On \textbf{sentiment}, the same shift produces $+6.2$ points for local champions and $+9.7$ for globals on a $[0,100]$ rescaling ($t = 2.52$, $p = 0.014$): a mild home-language warming for both tiers, with no reversal. A proprietary composite index (a Rankfor measure combining recommendation share, sentiment, and source quality, reported here only to corroborate the transparent measures) moves $+17.1$ for local champions versus $+2.5$ for globals ($t = -8.12$, $p < 0.001$), and the decomposition shows why: the gap lives in the recommendation-share component, not in sentiment.

The interpretation is direct. \emph{Query language changes which brands the model recommends far more than how it describes them.} A local champion is rarely named in English-language buyer-intent answers and is named in almost every home-language answer; a global brand is named at similar rates in either language. An English-only audit therefore understates a locally headquartered brand's AI visibility, while representing a multinational's fairly. This is the language blind spot, and it falls hardest on exactly the brands least able to afford English-language seeding.

\subsection{The Blind Spot, Quantified}\label{sec:blindspot}

On the main corpus, 45\% of the 65 brands show a home-language versus English sentiment gap above 0.15, with a mean absolute gap of 0.171. On the companion cohort, where each brand was queried across several local languages, 90\% of brands (18 of 20) show a divergence above 0.15 in at least one language; the mean absolute en-versus-local divergence is 0.287. The largest divergences are concentrated in Estonian (Wizz Air $+0.737$, Kiwi.com $+0.560$, CD Projekt $+0.557$, ESET $+0.555$), consistent with the home-language visibility effect for brands operating in smaller-language markets.

\subsection{Q1d: Language Clustering}\label{sec:clustering}

\begin{figure}[t]
\centering
\includegraphics[width=0.9\textwidth]{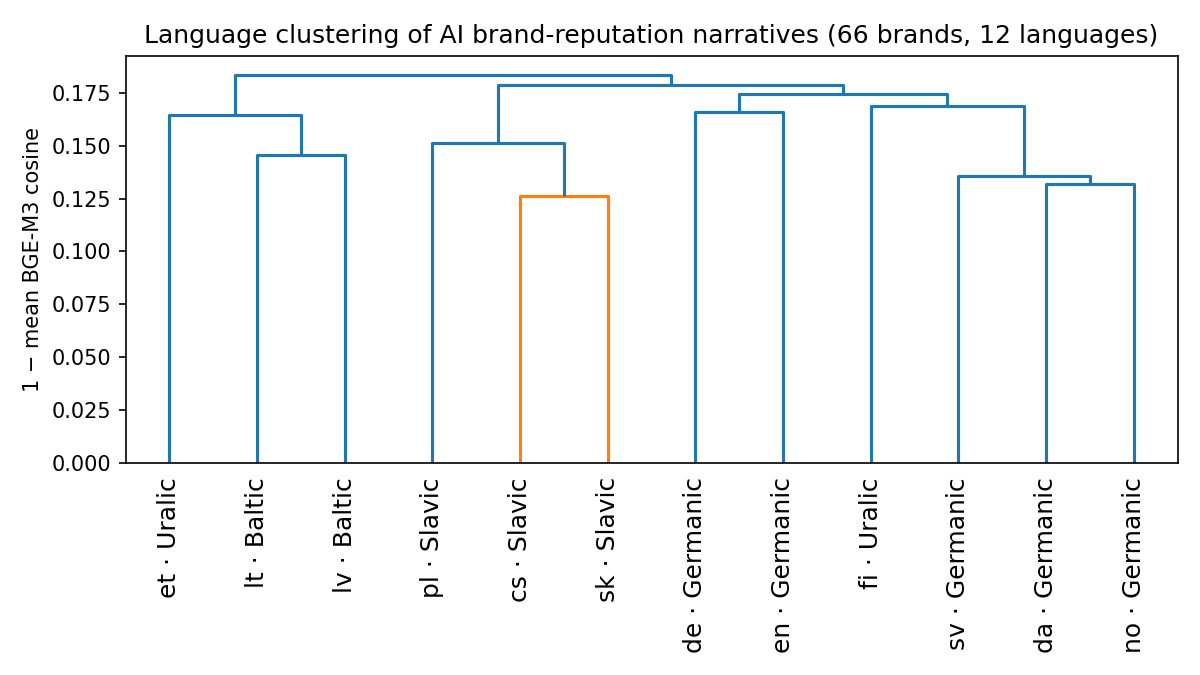}
\caption{Hierarchical clustering of 12 languages by AI-narrative similarity (average linkage on $1 - $ mean BGE-M3 cosine). Slavic $\{$cs, pl, sk$\}$ and Baltic $\{$et, lt, lv$\}$ recover as clean groups; Finnish clusters with the Germanic languages. Cophenetic correlation 0.915.}\label{fig:dendrogram}
\end{figure}

Average-linkage clustering of the 12 languages by mean AI-narrative similarity (Figure~\ref{fig:dendrogram}) has high dendrogram fidelity (cophenetic correlation 0.915). At a three-cluster cut the structure is interpretable: the Baltic languages and Estonian form one group $\{$et, lt, lv$\}$, the Slavic languages form a clean second group $\{$cs, pl, sk$\}$, and the remaining Germanic languages cluster with Finnish $\{$da, de, en, no, sv, fi$\}$. Language family thus predicts clustering, but imperfectly: the Slavic and Baltic families recover cleanly, while Finnish groups with the Germanic languages rather than with Estonian, plausibly because Finland's business and technology coverage leans heavily on English- and Swedish-language sources. Both structural language distance and shared information ecosystems appear to matter; neither alone accounts for the grouping.

\subsection{Q2: Stability Is Governed by Model, Not Language}\label{sec:stability}

On the companion cohort's five-iteration probes (1{,}435 stability cells), responses are highly repeatable: mean stability is 0.935 and 87.2\% of cells exceed 0.90. The variation that exists is explained more by model than by language: $\eta^2_{\text{model}} = 0.319$ against $\eta^2_{\text{language}} = 0.011$, a roughly thirty-fold difference. By model, Gemini (0.952) and OpenAI (0.950) are the most stable and Perplexity the least (0.904). By language the spread is narrow, with English highest (0.941) and Estonian lowest (0.925), echoing the same ordering reported on the earlier six-language cohort. The practical reading: for measurement reliability, which model you query matters far more than which language you query in. Stability is measured on the 20-brand companion cohort only, since the main corpus deferred the five-iteration extension; we treat it as a sub-study rather than a whole-corpus result.

\subsection{Source Structure}\label{sec:sources}

Source \emph{type} mix varies across languages by a statistically detectable but small margin (chi-square $\chi^2 = 1906.4$, $p < 0.001$, Cram\'{e}r's $V = 0.036$); at this sample size significance is easy and the effect size is the honest summary, so the type mix is broadly similar across languages. The source findings that hold concern citation concentration and specific domains, not the type mix. The citation base is long-tailed: 80\% of citations come from 3{,}878 of 21{,}077 domains (18.4\%), and a power-law fit on the top-ranked domains gives $\alpha \approx 0.85$ ($R^2 > 0.95$). Wikipedia is the single most-cited domain in 11 of the 12 languages; the exception is Lithuanian, where the national business daily \texttt{vz.lt} edges it out (4.38\% of Lithuanian citations). The marginal cases, where a national business outlet approaches Wikipedia, are concentrated in the smaller languages, consistent with localisation mattering at the margin.

\subsection{Robustness: How Many Dimensions Does AI Brand Visibility Have?}\label{sec:dimensionality}

An earlier exploratory study on 12 brands proposed that AI brand visibility reduces to two dimensions. We re-derive the structure here on 66 brands using standardised principal component analysis over six brand-level measures (sentiment, source quality, cross-language consistency, recommendation share, source concentration, citation count). The first two components account for 60.8\% of variance: a source-breadth axis (loading on source quality $+0.60$ against source concentration $-0.59$) and a prominence axis (loading on citation count $+0.65$, consistency $+0.55$, recommendation share $+0.42$). A Kaiser criterion, however, retains three components (eigenvalues 2.14, 1.56, 1.02). The two-dimensional account is therefore a useful approximation that captures most of the structure, but it does not stay clean at this larger scale; a third, weaker dimension emerges. We report dimensionality as approximate rather than settled, and caution against treating any two-factor summary of AI brand visibility as established.

\section{Discussion}\label{sec:discussion}

\subsection{What Is Said Is Language-Bound; How Stable It Is Is Model-Bound}

Two findings organise the results. The content of AI-constructed reputation, its semantic shape, its sentiment, and above all which brands it names, depends on query language (Sections~\ref{sec:semantic}--\ref{sec:clustering}). Its stability under repetition depends on model choice, not language (Section~\ref{sec:stability}). For a team deciding where to spend monitoring effort, the two findings point in different directions: cover multiple languages to capture variation in what is said, and cover multiple models to capture variation in how reliably it is said.

\subsection{The Blind Spot Falls on Local Champions}

The tier analysis (Section~\ref{sec:tierflip}) locates the practical risk precisely. The home-language effect is not mainly about tone; it is about visibility. Local champions move from near-absence in English buyer-intent answers to near-ubiquity in their home language, a 0.80 swing in recommendation share, while global brands barely move. An organisation auditing its AI presence only in English will read this as ``the model ignores us'' for a brand that is in fact the default recommendation at home, and will draw the opposite, falsely reassuring conclusion for a multinational whose home-market coverage is more critical. The asymmetry matters for resource allocation: the brands with the largest blind spot are the locally headquartered ones least likely to have invested in English-language presence.

\subsection{Measurement, Not Ground Truth}

We measure a property of model outputs, not human reputation. This boundary is deliberate. The outputs are consequential in their own right, because they are what users read, but we make no claim that the cross-language sentiment ordering or the recommendation tier-flip mirrors how people in each market actually regard these brands. Establishing that correspondence would require per-market human-perception baselines (for example RepTrak-style surveys), which we did not collect. The dimensionality re-derivation (Section~\ref{sec:dimensionality}) is offered in the same spirit: it tempers an earlier two-dimensional claim rather than confirming it.

\subsection{Limitations}\label{sec:limitations}

\emph{No human baseline.} The study measures AI output, not its correspondence to human perception. Without a per-market survey reference we cannot say whether cross-language divergence reflects real cross-cultural reputation differences or training-data artefacts.

\emph{Sentiment calibration.} XLM-RoBERTa scores long, structured responses as more negative than a human reader would (a clearly positive grounded answer can receive a negative score). We mitigate by leaning on translation-free embedding similarity and on relative orderings, but absolute sentiment magnitudes should be read with caution.

\emph{Stability is a sub-study.} The five-iteration stability data exist only for the 20-brand companion cohort; the main 66-brand corpus deferred that extension. The stability conclusions generalise to the main corpus only by assumption.

\emph{Response-level attribution.} Sentiment and source signals are scored at the response level. When a model lists several brands and cites one source, that source and sentiment attach to all brands named, which can inflate apparent ``bad-source'' associations. Entity-anchored scoring would tighten this.

\emph{Hand-classified recognition tier.} The global-versus-local split is expert-assigned. A quantitative derivation (from Wikipedia article length or web-index volume) would reduce subjectivity, though the tier contrast is large enough that modest misclassification is unlikely to overturn it.

\emph{Media-volume confound.} We did not include per-language media-coverage volume as a covariate, so we cannot separate language-family effects from coverage-density effects on sentiment.

\emph{Temporal and version scope.} Data were collected over a single window in April--May 2026 on specific model versions; grounded models change frequently, and the specific values are time- and version-dependent.

\section{Conclusion}\label{sec:conclusion}

Across 35{,}640 grounded responses about 66 brands in twelve languages and four language families, AI-constructed brand reputation is measurably language-bound. Responses diverge across languages (mean cross-language cosine 0.825), more within than across language families ($d = 0.31$), and sentiment varies systematically by language ($\eta^2 = 0.077$), with Uralic and Baltic languages the most positive and Germanic languages the most critical. Clustering the twelve languages recovers the Slavic and Baltic families cleanly (cophenetic 0.915). The most consequential effect is on visibility rather than tone: querying in a brand's home language raises recommendation share by 0.80 for local champions but only 0.15 for global multinationals, so English-only auditing systematically understates locally headquartered brands. Stability under repeated querying, by contrast, is governed by model choice, not language. An earlier two-dimensional account of AI brand visibility holds only approximately at this scale.

The measurements indicate that the assumption behind most AI-visibility monitoring, that an English-language query returns a representative picture, does not hold for brands operating in multilingual European markets. The language blind spot is real, it is concentrated in recommendation rather than sentiment, and it falls hardest on local champions. These results connect to corporate reputation management, AI fairness \cite{ekstrand2019fairness}, and generative engine optimisation \cite{aggarwal2023geo, shin2025hbr}, and they motivate multilingual, multi-model auditing as the default rather than the exception.

\backmatter
\bookmarksetup{startatroot}

\bmhead{Data Availability}

The dataset (raw responses, per-response sentiment and embeddings, citation attributions, per-(brand $\times$ language) measures), the analysis notebook, and the prompt templates in all twelve languages are deposited on Zenodo at \url{https://doi.org/10.5281/zenodo.20794390} (CC BY 4.0). A live, browsable version of the underlying index is available at \url{https://open.rankfor.ai/index-2026}.

\bmhead{Code Availability}

The data-collection and analysis pipeline is provided in the Zenodo deposit. The five-iteration dice-roll collection infrastructure uses the open-source Dice Roller package at \url{https://github.com/Rankfor/rankfor-open/tree/main/dice-roller}.

\bmhead{Acknowledgements}

The author thanks the Estonian Entrepreneurship University of Applied Sciences for institutional support.

\section*{Declarations}

\begin{itemize}
\item \textbf{Funding:} This research received no external funding.
\item \textbf{Competing interests:} D.~\.Zatuchin is affiliated with Rankfor.AI, which develops AI brand-intelligence tools. The research was conducted independently; the company had no influence on study design, methodology, analysis, or conclusions. The composite index referenced in Section~\ref{sec:tierflip} is a commercial product of Rankfor.AI and is reported only as corroboration of the transparent, model-free measures.
\item \textbf{Ethics approval:} Not applicable. The study analysed publicly accessible AI systems and did not involve human subjects.
\item \textbf{Consent to participate / for publication:} Not applicable.
\item \textbf{Author contribution:} D.\.Z.\ conceived the study, designed the methodology, implemented the collection infrastructure, performed the analysis, and wrote the manuscript.
\end{itemize}

\begin{appendices}

\section{Sentiment by Language and Model}\label{secA1}
\begin{table}[htbp]
\caption{Mean sentiment by model (across all languages), main corpus.}\label{tab:sentiment_matrix}
\begin{tabular*}{\textwidth}{@{\extracolsep\fill}lc}
\toprule
\textbf{Model} & \textbf{Mean sentiment} \\
\midrule
OpenAI GPT-5.4 & $+0.074$ \\
Perplexity Sonar Pro & $-0.003$ \\
Gemini 3.1 Pro & $-0.016$ \\
\bottomrule
\end{tabular*}
\end{table}

\section{Stability by Model (Companion Cohort)}\label{secA2}
\begin{table}[htbp]
\caption{Mean five-iteration stability by model, companion cohort. Higher is more repeatable.}\label{tab:stability_model}
\begin{tabular*}{\textwidth}{@{\extracolsep\fill}lc}
\toprule
\textbf{Model} & \textbf{Mean stability} \\
\midrule
Gemini & 0.952 \\
OpenAI & 0.950 \\
Perplexity & 0.904 \\
\bottomrule
\end{tabular*}
\end{table}

\section{Dimensionality (PCA) Loadings}\label{secA3}
\begin{table}[htbp]
\caption{First two principal components of brand-level AI-visibility measures (66 brands, standardised). PC1+PC2 account for 60.8\% of variance; a Kaiser criterion retains three components.}\label{tab:pca}
\begin{tabular*}{\textwidth}{@{\extracolsep\fill}lcc}
\toprule
\textbf{Measure} & \textbf{PC1 (source breadth)} & \textbf{PC2 (prominence)} \\
\midrule
Source quality & $+0.60$ & $-0.17$ \\
Source concentration (HHI) & $-0.59$ & $+0.22$ \\
Cross-language consistency & $+0.30$ & $+0.55$ \\
Recommendation share & $+0.38$ & $+0.42$ \\
Citation count & $-0.10$ & $+0.65$ \\
Sentiment & $-0.23$ & $+0.13$ \\
\midrule
\textbf{Variance explained} & 35.2\% & 25.7\% \\
\bottomrule
\end{tabular*}
\end{table}

\end{appendices}


\end{document}